%% file: Journal_Main.tex
\documentclass[Journal]{IEEEtran}
\IEEEoverridecommandlockouts
\usepackage{cite}
\usepackage{amsmath,amssymb,amsfonts}

\usepackage{textcomp}
\usepackage{xcolor}
\usepackage{multirow}
\usepackage{makecell}
\usepackage{algorithm,algcompatible}
\usepackage{subfigure}
\usepackage{graphicx}
\usepackage{array}
\usepackage{subcaption}
\newcolumntype{P}[1]{>{\centering\arraybackslash}p{#1}}

\usepackage{booktabs}

\def\BibTeX{{\rm B\kern-.05em{\sc i\kern-.025em b}\kern-.08em
    T\kern-.1667em\lower.7ex\hbox{E}\kern-.125emX}}

\input{macros}

\DeclareMathOperator*{\argmax}{arg\,max}

\newtheorem{remark}{Remark}

\newtheorem{definition}{Definition}

\begin{document}


\title{Near-Field LoS/NLoS Channel Estimation for RIS-Aided MU-MIMO Systems: Piece-Wise Low-Rank Approximation Approach}



\author{Jeongjae~Lee,~\IEEEmembership{Student Member,~IEEE}
        and~Songnam~Hong,~\IEEEmembership{Member,~IEEE}

\thanks{This work was supported in part by the Technology Innovation Program (1415178807, Development of Industrial Intelligent Technology for Manufacturing, Process, and Logistics) funded By the Ministry of Trade, Industry \& Energy(MOTIE, Korea) and in part by the Institute of Information \& communications Technology Planning \& Evaluation (IITP) under the artificial intelligence semiconductor support program to nurture the best talents (IITP-2024-RS-2023-00253914) grant funded by the Korea government(MSIT).}}

\maketitle




\begin{abstract}
We study the channel estimation problem for a reconfigurable intelligent surface (RIS)-assisted millimeter-wave (mmWave) multi-user multiple-input multiple-output (MU-MIMO) system. In particular, it is assumed that the channel between a RIS and a base station (BS) exhibits a near-field line-of-sight (LoS) channel, which is a dominant signal path in mmWave communication systems. Due to the high-rankness and non-sparsity of the RIS-BS channel matrix in our system, the state-of-the-art (SOTA) methods, which are constructed based on far-field or near-field non-LoS (NLoS) channel, cannot provide attractive estimation performances. We for the first time propose an efficient near-field LoS/NLoS channel estimation method for RIS-assisted MU-MIMO systems by means of a piece-wise low-rank approximation. Specifically, an effective channel (to be estimated) is partitioned into piece-wise effective channels containing low-rank structures and then, they are estimated via collaborative low-rank approximation.  The proposed method is named PW-CLRA. Via simulations, we verify the effectiveness of the proposed PW-CLRA.
\end{abstract}

\begin{IEEEkeywords}
Reconfigurable intelligent surface (RIS), channel estimation, hybrid beamforming, piece-wise near-field channel.
\end{IEEEkeywords}

%
\section{Introduction}\label{sec:Intro}

Millimeter-wave (mmWave) communication systems, which operate at higher frequencies (e.g., 30 - 300 GHz) and larger bandwidths, can provide higher data rates \cite{wang2018millimeter}. In these systems, however, the signals become susceptible to blockages due to the severe path-loss and high directionality. Reconfigurable intelligent surface (RIS) is an emerging technique for robust mmWave multiple-input multiple-output (MIMO) systems \cite{di2020smart, pei2021ris, wu2021intelligent}. A RIS is composed of a uniform array with a large number of reflection elements, which can create a favorable propagation environment by controlling the phase and reflection angle of an incident signal \cite{long2020active}. In the deployment of RIS-assisted MIMO systems, it is typically challenging to acquire a channel state information (CSI) accurately, thereby making it hard to fully exploit the advantages of the RIS.


There have been numerous works on the design of an efficient channel estimation method for RIS-assisted mmWave MIMO systems. In these works, the channels between the RIS and the base station (BS) (in short, the RIS-BS channel), and between the users and the RIS (in short, the User-RIS channels) are commonly modeled as {\em far-field} channels with {\em planar-wavefront} assumption. We remark that in this modeling, the channels can be represented by the product of far-field array response vectors (i.e., decomposability). Since the RIS is not capable of a signal processing, the channel estimation problem of the RIS-assisted systems aims at acquiring the so-called {\em effective channel} (a.k.a. the cascaded channel). Regarding the far-field channel estimations, a compressed sensing (CS)-based methods were presented to improve an estimation accuracy while having an affordable computational complexity \cite{tsai2018efficient, chen2023channel}. These methods were developed by harnessing the sparsity and decomposability of the far-field mmWave channels.
However, the CS-based methods suffer from an inevitable {\em grid-mismatch} problem \cite{chi2011sensitivity}, which is caused by the quantization error in the construction of a dictionary matrix by quantizing the array response vectors at a certain resolution. In \cite{Lee2024near, chung2023location}, it was shown that having an affordable computational complexity (i.e., a coarse quantization), the CS-based methods result in a severe error-floor problem. In \cite{He2021, chung2023location}, the gird-mismatch problem was alleviated via an atomic norm minimization (ANM). Yet, 
the ANM-based method is impractical due to an expensive computational complexity.  Recently in \cite{JJLee2023, chung2023location}, the channel estimation problem was formulated as a low-rank matrix completion (LRMC) and then solved via a fast alternating least squares (FALS) \cite{hastie2015matrix}. Also, it was shown that the LRMC-based method can outperform the CS-based methods by avoiding the grid-mismatch problem \cite{chung2024efficient, JJLee2023}. However, it suffers from the {\em noisy-sample} problem as the sampled noisy elements cannot be updated during the matrix completion process.


With the advancement of an antenna technology, an extremely large-scale MIMO (XL-MIMO) has been proposed as a feasible and promising candidate to achieve a higher data rate \cite{cui2022,Lu2023}. As the number of antennas considerably increases in the XL-MIMO systems (e.g., 1020 antennas \cite{zhu2021bayesian}), the near-field range will expand by orders of magnitude, which can be up to several hundreds of meters \cite{Lu2023}. Thus, the XL-MIMO channel should be modeled by the {\em near-field} channel with {\em spherical-wavefront} assumption. Similarly to the far-field channel, the near-field channel was modeled as the product of {\em near-field} array response vectors \cite{Yu2023, yang2023, Yang2024,Lee2024near}. Due to the spherical wave assumption, they relate to the distance as well as the angle, whereas the far-field array response vectors only rely on the angle. In \cite{yang2023,Yang2024}, utilizing the polar-domain sparsity, a CS-based channel estimation method was proposed for the near-field RIS-assisted MU-MIMO systems. Herein, the dictionary matrix is totally different from those in the CS-based far-field counterparts in \cite{tsai2018efficient, chen2023channel} since they should be carefully designed according to the structures of the array response vectors. Recently in \cite{Lee2024near}, an efficient channel estimation method (named CLRA) was proposed for RIS-assisted MU-MIMO systems with hybrid beamforming structures. This method was constructed by  harnessing the low-rankness of the RIS-BS channel. Noticeably, the CLRA can be applied to the both far-field and near-field channels without any modification, whereas the CS-based methods require a suitable design of the dictionary matrix according to the categories of wireless channels. Also, it was demonstrated that the CLRA can achieve a higher estimation accuracy in the far-field, near-field, and mixed far-field/near-field channels than the corresponding CS-based methods while having a lower training overhead. Therefore, the CLRA would be a good candidate for typical mixed near-field/far-field RIS-assisted MU-MIMO systems.

Very recently, it was claimed in \cite{Lu2023} that the near-field channel modeling in the above can accurately describe the non-line-of-sight (NLoS) path components while leading to inaccurate description of the line-of-sight (LoS) path component. To be specific, the near-field LoS path components cannot be expressed as the product of array response vectors if the distance between the transmitter and the receiver is shorter than a certain value. This value was characterized in \cite{Lu2023}, which is referred to as MIMO advanced Rayleigh distance (MIMO-ARD). As the carrier frequency becomes higher (e.g., mmWave and THz channels) and the number of antennas grows, the corresponding channel is likely to belong to the near-field LoS region. Motivated by this, a more realistic {\em mixed LoS/NLoS near-field channel} was provided in \cite{Lu2023}. In general, this mixed channel does not contain the structures of the sparsity and low-rankness. Thus, the aforementioned CS-based methods and CLRA suffer from a performance loss in the mixed LoS/NLoS near-field XL-RIS aided XL-MIMO systems. To the best of our knowledge, there is no efficient channel estimation method for this mixed near-field channel.


In this paper, we investigate a mixed LoS/NLoS near-field channel estimation for RIS-assisted mmWave MU-MIMO system. For practicality, moreover, a hybrid beamforming structure at the BS and multi-user scenario are considered. In this system, we for the first time propose an efficient channel estimation method by means of a piece-wise low-rank approximation. The proposed method is referred to as {\bf P}iece-{\bf W}ise {\bf C}ollaborative {\bf L}ow-{\bf R}ank {\bf A}pproximation (PW-CLRA). Remarkably, without any modification, the proposed PW-CLRA can be used in any uniform array structure (e.g., (e.g., uniform linear array (ULA), uniform planar array (UPA), etc.). Toward this, our major contributions are summarized as follows.
\begin{itemize}
    \item For the mixed LoS/NLoS near-field channel estimation, the major challenge is that the $K$ effective channels (to be estimated) do not contain the sparse and low-rank structures, where $K$ denotes the a number of users in the system. In fact, these are the key ingredients to develop the aforementioned state-of-the-art (SOTA) methods.
    
    \item In the proposed PW-CLRA, each effective channel is partitioned into $Q$ piece-wise effective channels, where $Q\geq 1$ is a design parameter. We identify that the $K$ corresponding piece-wise effective channels share a low-rank subspace. Leveraging this, the proposed method estimates the $K$ piece-wise effective channels jointly but in parallel with respect to each piece.

    \item In the first phase of the training, the shared subspace is estimated using the $K$ users' training observations. To enable the piece-wise estimation, the reflection vector at the RIS and the RF combining matrix at the BS are carefully designed, which are completely different from those in the far-field (or NLoS-only near-field) counterpart \cite{Lee2024near}.

   \item In the second phase of the training, harnessing the special property of the $K$ piece-wise effective channels, the user-specific coefficient matrices are jointly optimized. The superiority of the proposed joint optimization compared with the individual minimum mean-square error (MMSE) estimation is also verified. From the estimated shared subspace and the user-specific coefficient matrices, the $K$ effective channels are straightforwardly obtained.
 
    \item Via simulations, we verify the effectiveness of the proposed PW-CLRA in various channel environments. In comparison with the SOTA method under the fair conditions, it is shown that the proposed method can reduce the training overhead by $70\%$ while achieving target accuracy. Furthermore, we demonstrate that the proposed PW-CLRA can provide the stable performances to the increased number of RIS reflection elements, thereby being suitable for large-scale RIS-assisted systems. Noticeably, it is verified that the proposed method can be applied to the various uniform antenna arrays without any modification. Due to its attractive performance, scalability, and flexibility, we can conclude that the proposed PW-CLRA would be a good candidate for the near-field channel estimation in RIS-assisted MU-MIMO systems.
    

    
\end{itemize}


The remaining part of this paper is organized as follows. In Section II, we describe the signal and channel models for the RIS-assisted MU-MIMO systems. Section III introduces the channel estimation protocol and the frame structures. The proposed PW-CLRA is described in Section IV. Section V provides simulation results and Section VI concludes the paper.

{\em Notations.} Let $[N]\eqdef \{1,2,...,N\}$ for any positive integer $N$.
We use $\xv$ and $\Am$ to denote a column vector and matrix, respectively. Also, $\Am^{\dagger}$ denotes the Moore-Penrose inverse and $\otimes$ denotes the Kronecker product. Given a $M \times N$ matrix $\Am$, let $\Am(i,:)$ and $\Am(:,j)$ be the $i$-th row and $j$-th column of $\Am$, respectively, and $\|\Am\|_F$ denotes the Frobenius norm of $\Am$. Also, given the index subsets $\Ic_{\rm row}\subseteq [M]$ and $\Ic_{\rm col} \subseteq [N]$,  let $\Am(\Ic_{\rm row},:)$ and $\Am(:,\Ic_{\rm col})$ be the submatrices of $\Am$ by only taking the rows and columns whose indices are belong to $\Ic_{\rm row}$ and $\Ic_{\rm col}$, respectively. Given a vector $\vv$, ${\rm diag}(\vv)$ denotes a diagonal matrix whose $\ell$-th diagonal element is equal to the $\ell$-th element of $\vv$. We let $\Id_M$, ${\bf 0}_M$ and ${\bf 1}_M$ denote the $M\times M$ identity and all-zero matrices and $M\times 1$ all-one vector, respectively.

\begin{figure}[t]
\centering
\includegraphics[width=1.0\linewidth]{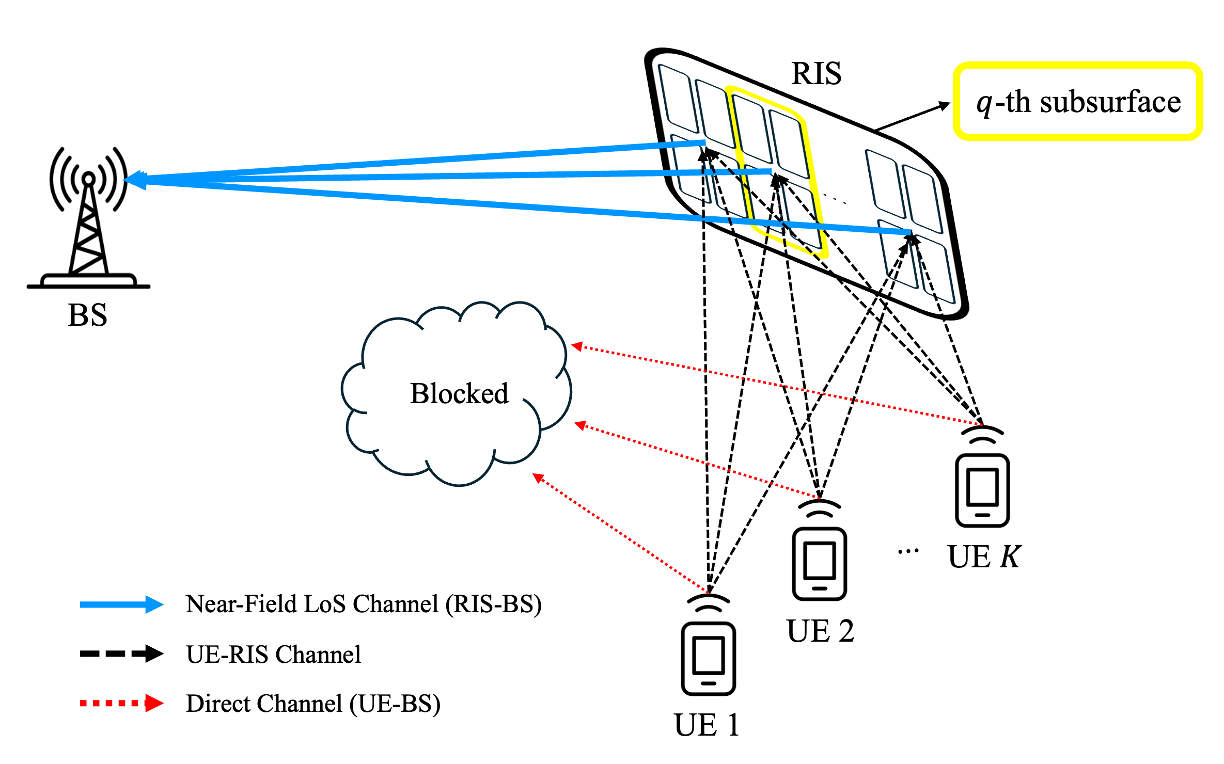}
\caption{The description of the RIS-aided MU-MIMO system.}
\end{figure}

\section{System Model}\label{sec:System Model}

In this section, we describe the signal model of the RIS-assisted MU-MIMO system and specify the wireless channels of the RIS-BS and User-RIS channels.

\subsection{Signal Model}

We consider an uplink time-division-duplexing (TDD) multi-user multiple-input multiple-output (MU-MIMO) system, in which a base station (BS) with $N$ transmit antennas serves the $K$ users each having $L$ receiver antennas. For the sake of lower complexity, cost, and power consumption \cite{molisch2017hybrid}, it is assumed that the BS uses the hybrid beamforming structure sharing a limited number of radio frequency (RF) chains among BS antennas, where $N_{\rm RF}$ denotes the number of RF chains at the BS. Throughout the paper, $B=N/N_{\rm RF}$ is assumed to be an integer. As shown in Fig. 1, a reconfigurable intelligent surface (RIS) is deployed for robust millimeter-wave (mmWave) MU-MIMO systems, where the RIS is equipped with $M$ reflective elements. The reflection vector in the RIS is defined as 
\begin{equation}
    \vv = \left[v_{1},v_{2},\dots,v_{M}\right]^{\intercal},
\end{equation} with $v_{m} = \beta_{m}e^{j\vartheta_{m}}$. We highlight that the proposed channel estimation method can be applied to any type of RIS, i.e., metasurface \cite{Ayanoglu2022} or discrete phase shifter \cite{Wu2020} without any modification. For the low-complexity of hardware implementations, the letter one is commonly considered in the current literature \cite{Wu2020,Moon2022}, where $|\beta_m|=1$, $\vartheta_{m}\in\Theta$ for  $\forall m\in[M]$, and $\Theta$ is a finite set of possible phase elements.
As considered in the related works\cite{Lee2024near,yang2023,Yang2024}, the direct channels between the BS and the $K$ users are assumed to be blocked due to the several obstacles and the severe path-loss of mmWave communications. The channel responses from the user $k$ to the RIS and from the RIS to the BS are denoted by $\Hm_{k}^{\rm UR}\in\CC^{M\times L}$ and $\Hm^{\rm RB}\in\CC^{N\times M}$, respectively. Given a reflection vector $\vv$, the total cascaded channel response from the user $k$ to the BS is defined as
\begin{align}
    \Hm_{k}^{\rm tot}\left(\vv\right) &\eqdef \Hm^{\rm RB}{\rm diag}\left(\vv\right)\Hm_{k}^{\rm UR}\nonumber\\
    &=\Hm_{k}^{\rm eff}\times \left(\Id_{M}\otimes\vv\right), \label{eq:cascadedchannel}
\end{align} where 
\begin{equation}
    \Hm_k^{\rm eff}=\left[\Hm_{[k,1]}^{\rm eff} \cdots \Hm_{[k,L]}^{\rm eff}\right] \in\CC^{N\times ML}.
\end{equation} is called an effective (or cascaded) channel. Herein, $\Hm_{[k,\ell]}^{\rm eff}$ (called a column-wise effective channel) is expressed as
\begin{equation}
     \Hm_{[k,\ell]}^{\rm eff}\eqdef\Hm^{\rm RB}\mbox{diag}\left(\Hm_k^{\rm UR}(:,\ell)\right)\in\CC^{N\times M}.\label{eq:CW}
\end{equation} For the joint optimization of a reflection vector and hybrid beamforming matrices, it is necessary to acquire the effective channels $\{\Hm_{k}^{\rm eff}: k\in [K]\}$ \cite{Moon2022,Choi2024}. Motivated by this, we aim at developing an efficient method to estimate the effective channel (equivalently, the column-wise effective channels).

\subsection{Channel Model}

We describe the RIS-BS channel (i.e., $\Hm^{\rm RB}$). Beyond the well-studied far-field channels \cite{chen2023channel,He2021}, we consider a near-field XL-MIMO channel, which is composed of line-of-sight (LoS) and non-LoS path components. It is called a {\em mixed NLoS/LoS near-field channel} \cite{Lu2023}. The corresponding channel matrix is represented as
\begin{equation}
    \Hm^{\rm RB} = \Hm^{\rm RB}_{\rm LoS} + \Hm^{\rm RB}_{\rm NLoS},
\end{equation} where $\Hm^{\rm RB}_{\rm LoS}$ and $\Hm^{\rm RB}_{\rm NLoS}$ represent the near-field LoS and NLoS channel matrices, respectively. They will be specified in the below.

As identified in \cite{Lu2023}, the near-field LoS channel is experienced when the distance between the RIS and the BS (denoted by $d^{\rm RB}$) is shorter than  the advanced MIMO Rayleigh distance (MIMO-ARD) (denoted by $D^{\rm RB}$), namely, 
$ d^{\rm RB}<D^{\rm RB}$.
From \cite{Lu2023}, the MIMO-ARD of the RIS-BS channel is given by
\begin{equation}
    D^{\rm RB} =  \lambda NM,
\end{equation} where $\lambda$ is the wavelength of the carrier signals. In the near-field LoS channel, all the element-wise distances between the RIS reflective elements and the BS antenna components should be considered \cite{Lu2023}:
\begin{equation}
    \Hm_{\rm LoS}^{\rm RB} = \begin{bmatrix}
    \bv_{1}^{\rm RB} &\cdots& \bv_{M}^{\rm RB}
    \end{bmatrix}\in\CC^{N\times M},\label{eq:LoS1}
\end{equation} where, a near-field array response vector from the $m$-th reflective element of the RIS to the BS is determined by
\begin{equation}
    \bv_{m}^{\rm RB}\eqdef\left[\alpha_{[m,1]}^{\rm RB}e^{-j\frac{2\pi}{\lambda}r_{[m,1]}^{\rm RB}}, \dots, \alpha_{[m,N]}^{\rm RB}e^{-j\frac{2\pi}{\lambda}r_{[m,N]}^{\rm RB}}\right]^{\intercal}.\label{eq:RBresponse}
\end{equation}  Herein,  $\alpha_{[m,n]}^{\rm RB}\eqdef\frac{\lambda^2}{\left(4\pi r_{[m,n]}^{\rm RB}\right)^2}$ is the associated path coefficient (i.e., free space path-loss) and $r_{[m,n]}^{\rm RB}$ is the physical distance between the $m$-th reflective element of the RIS and the $n$-th antenna of the BS. Next, the near-field NLoS channel $\Hm_{\rm NLoS}^{\rm RB}$ can be represented as
\begin{equation} 
    \Hm_{\rm NLoS}^{\rm RB} = \sum_{p=1}^{N_{\rm NLoS}^{\rm RB}}\av_{[{\rm r},p]}^{\rm RB}\left(\av_{[{\rm t},p]}^{\rm RB}\right)^{\herm},
\end{equation} where $N_{\rm NLoS}^{\rm RB}$ denotes the number of NLoS paths, and the near-field array response vectors from the $p$-th scatter to the BS and from the RIS to the $p$-th scatter are respectively denoted as $\av_{[{\rm r},p]}^{\rm RB}$ and $\av_{[{\rm t},p]}^{\rm RB}$. And, these array response vectors are identically defined as the vector in \eqref{eq:RBresponse} considering the distance from the scatters and system wavelength. In \cite{Lu2023,cui2022,Lee2024near}, it is well-known that the near-field NLoS channel exhibits low-rank and sparsity, which are the key factors to construct efficient channel estimation methods by means of low-rank approximation and compressed sensing, respectively. However, in our mixed channel model, such properties cannot be utilized because $\Hm^{\rm RB}$ exhibits high rank and non-sparsity due to the LoS path component $\Hm_{\rm LoS}^{\rm RB}$ in \eqref{eq:LoS1}.

\vspace{0.1cm}
\begin{remark}
For ease of exposition, in this paper, we describe the proposed channel estimation method by assuming that the BS, RIS, and $K$ users are equipped with a uniform liner array (ULA). We emphasize that the proposed method can be applied to other uniform arrays (e.g., uniform planar array (UPA)) without any modification. This extension will be verified in Section~\ref{sec:SR}. Also, it is assumed that the User-RIS channels (denoted by $\{\Hm_{k}^{\rm UR}:k\in[K]\}$) can be any category of wireless channel (i.e., either a near-field or a far-field channel). Noticeably, the proposed channel estimation method in Section IV can operate regardless of the characteristics of the User-RIS channels. This advantage makes the proposed method more practical as the User-RIS channels of moving users can be rapidly varied.
\hfill$\blacksquare$
\end{remark}

%
\section{Channel Estimation Protocol}\label{sec:CEP}

The proposed channel estimation protocol proceeds with $Z$ subframes, each of which consists of $T$ symbols (or time slots) with $T \geq KL$. Throughout the paper, $Z$ is referred to as the training overhead. 
For each subframe $z \in [Z]$, the reflection vector $\vv_z$ and the $N\times N_{\rm RF}$ RF combining matrix $\Cm_z$, which have the constant modulus constraints \cite{Lee2024asym}, are unchanged while they can be changed across the subframes. Then, $\{\vv_z, \Cm_z: z\in [Z]\}$ will be carefully designed such that the observations at the BS are suitable for estimating the effective channels (see Section~\ref{sec:method} for details).

During the $Z$ subframes, our channel estimation protocol proceeds in the following ways. For every $z \in [Z]$, each user $k \in [K]$ transmits its orthogonal pilot sequence of the length $T$ to the BS, denoted by $\Xm_{[z,k]} \in \CC^{L\times T}$, where
\begin{equation}
    \Xm_{[z,k]}\Xm_{[z,k]}^{\herm} =PT\times \Id_{L}\; \mbox{and}\; \Xm_{[z,k]}\Xm_{[z,k']}^{\herm} = {\bf 0}_L,\label{eq:power}
\end{equation} if $k\neq k'$, and $P$ denotes the transmit power per antenna. Note that each user $k$ sends the $\ell$-th row of $\Xm_{[z,k]}$ through the $\ell$-th transmit antenna during $T$ time slots. In the subframe $z \in [Z]$, the BS can observe the $N_{\rm RF} \times T$ matrix as
\begin{align}
    \Ym_z &= \Cm_{z}^{\herm}\sum_{k=1}^{K}\Hm_k^{\rm tot}\left(\vv_z\right)\Xm_{[z,k]} + \Cm_{z}^{\herm}\Um_z,
\end{align} where $\Um_{z}$ denotes the noise matrix whose elements follow independently identically circularly symmetric complex Gaussian distribution with mean zero and variance $\sigma^2$.

Using the orthogonality of the pilot sequences, the BS can derive the $N_{\rm RF}\times L$ user-specific observation matrix:
\begin{align}
\Ym_{[z,k]} &= \frac{1}{PT}\Ym_{z}\Xm_{[z,k]}^{\herm}\nonumber\\
&=\Cm_{z}^{\herm}\Hm_k^{\rm tot}\left(\vv_z\right) + \Cm_{z}^{\herm}\Um_{[z,k]},
\label{eq:variance}
\end{align} where $\Um_{[z,k]}\eqdef\frac{1}{PT}\Um_z\Xm_{[z,k]}^{\herm}$. In Section~\ref{sec:method}, we will estimate the effective channels $\{\Hm_{[k,\ell]}^{\rm eff}: k \in [K]\}$ in \eqref{eq:CW} from the processed observations $\{\Ym_{[z,k]}: k\in[K], z\in[Z]\}$.

\section{The Proposed PW-CLRA}\label{sec:method}

We investigate a channel estimation for RIS-assisted MU-MIMO systems especially when the RIS-BS channel exhibits a near-field LoS/NLoS channel. In this work, it is demanding to handle  the near-field LoS channel (defined in \eqref{eq:LoS1}) due to its non-sparsity and high-rankness. Since this channel cannot be decoupled by the near-field array response vectors, it is impossible to employ a polar domain sparsity. Thus, the popular compressed sensing-based channel estimation method in \cite{cui2022,Lu2023,Yang2024} cannot be adopted. The low-rank approximation-based method (named CLRA), which was recently proposed in \cite{Lee2024near}, can be applied to the near-field LoS channel. Due to the high-rankness of the RIS-BS channel, however, the CLRA requires a expensive training overhead. Specifically, in \cite{Lee2024near}, the training overhead of the CLRA is given by
\begin{equation}
    Z_{\rm CLRA}=B_c(N/N_{\rm RF}) + B_r M,\label{eq:CLRA_JO}
\end{equation} where the hyperparameters $B_c$ and $B_r$ can be chosen under the following conditions:
\begin{align}
B_c &\geq \frac{M}{KL}\; \mbox{and} \;B_r \geq \frac{\hat{\mbox{rk}}(\Hm^{\rm RB})}{ N_{\rm RF}},\label{CLRA_JO:constraint}
\end{align} where $\hat{\mbox{rk}}(\Hm^{\rm RB})$ denotes the estimated (or approximated) rank of the RIS-BS channel via MDL \cite{Lee2024near}. We can identify that the training overhead of the CLRA tends to increase as the rank of the RIS-BS channel (i.e., $\hat{\mbox{rk}}(\Hm^{\rm RB})$) grows (see Remark~\ref{remark:overhead} for details). Therefore, the CLRA is not suitable for the near-field LoS channel estimation considered in this paper. Motivated by this, we will develop an efficient channel estimation method suitable for the near-field LoS channels by means of a piece-wise low-rank matrix approximation.


\subsection{Piece-Wise Low-Rank Matrix Approximation}\label{subsec:PWLRMA}

We provide the fundamental idea of the proposed channel estimation method. To relax the rank-constraint in \eqref{CLRA_JO:constraint}, we divide each effective channel $\Hm_{[k,\ell]}^{\rm eff}$ into the $Q$ piece-wise effective channels (denoted by $\{\Hm_{[k,\ell,q]}^{\rm eff}: q \in [Q]\}$), each of which can be well-approximated as a low-rank matrix. To accomplish this, we partition the RIS surface into the $Q$ subsurfaces as shown in Fig. 1. Our intuition of this partition is as follows. The array response vectors associated with each subsurface have high-correlation as the distance discrepancy (or the phase discrepancy) within the subsurface is relatively small. Namely, the corresponding channel matrix between the subsurface of the RIS and the BS can be well-approximated as a low-rank matrix.

We explain the main idea focusing on the subsurface $q \in [Q]$. The exactly same procedures will be applied to the other subsurfaces in parallel. Let $\Mc_{q}\eqdef \{1+M_{\rm sub}(q-1),..., M_{\rm sub}q\}$ be the index subset of the columns of $\Hm^{\rm RB}$, where $M_{\rm sub}=|\Mc_{q}|$ is the number of the RIS reflection elements in each subsurface with $M=M_{\rm sub}Q$. In this subsurface, the corresponding submatrix of the RIS-BS channel (i.e., $\Hm_{q}^{\rm RB}\eqdef\Hm^{\rm RB}(:,\Mc_q)$) can be represented such as
\begin{align}
    \Hm_{q}^{\rm RB}&=\Sm_q^{\rm RB}\Tm_{q}^{\rm RB},\; q\in [Q], \label{eq:lowrank}
\end{align}  where $\Sm_q^{\rm RB}$ is the $N\times {\rm rk}\left(\Hm^{\rm RB}_{q}\right)$ matrix whose columns are
the basis of the column space of $\Hm^{\rm RB}_{q}$ (a.k.a. the piece-wise common column space), ${\rm rk}\left(\Hm^{\rm RB}_{q}\right)\leq\min\left(N,M_{\rm sub}\right)$ is the rank of $\Hm^{\rm RB}_{q}$, and $\Tm_q^{\rm RB}\in\CC^{{\rm rk}\left(\Hm^{\rm RB}_{q}\right)\times M_{\rm sub}}$ is the piece-wise coefficient matrix. From the the column-wise effective channel in \eqref{eq:CW}, we define the piece-wise effective channel, which is only associated with the subsurface $q$:
\begin{equation}
    \Hm_{[k,\ell,q]}^{\rm eff}\eqdef\Hm_{[k,\ell]}^{\rm eff}(:,\Mc_q) \in \CC^{N\times M_{\rm sub}}.
\end{equation} From \eqref{eq:CW} and \eqref{eq:lowrank}, it can be expressed as
\begin{equation}
    \Hm_{[k,\ell,q]}^{\rm eff} = \Sm_q^{\rm RB}\Tm_{[k,\ell,q]},\label{eq:qeff}
\end{equation} where $\Dm_{[k,\ell,q]} \eqdef {\rm diag}\left(\Hm_{k}^{\rm UR}(:,\ell)\right)$ and
\begin{equation}
    \Tm_{[k,\ell,q]} = \Tm_{q}^{\rm RB}\Dm_{[k,\ell]}(\Mc_q,\Mc_q).\label{eq:Tklq}
\end{equation} This is due to the following fact:
\begin{align}
    \Hm_{[k,\ell]}^{\rm eff}&=\Hm^{\rm RB}\rm{diag}\left(\Hm_{k}^{\rm UR}(:,\ell)\right)\nonumber\\
    &= \sum_{q=1}^{Q}\Hm_{q}^{\rm RB}\Dm_{[k,\ell]}(\Mc_q,:)\nonumber\\
    &=\sum_{q=1}^{Q}\Sm_{q}^{\rm RB}\Tm_{q}^{\rm RB}\Dm_{[k,\ell]}(\Mc_q,:),\label{eq:AppLM}
\end{align} and $\Tm_{q}^{\rm RB}\Dm_{[k,\ell]}(\Mc_q,:)$ has only non-zero columns whose indices belong to $\Mc_q$. From \eqref{eq:Tklq}, the coefficient matrices $\{\Tm_{[k,\ell,q]}: \ell \in [L], k \in [K]\}$ satisfy the so-called {\em scaling property} such as
\begin{equation}
    \Tm_{[k,\ell,q]} = \Tm_{[1,1,q]}\Dm_{[1,1]}(\Mc_q,\Mc_q)\Dm'_{[k,\ell,q]},\label{eq:scaling}
\end{equation} where the diagonal matrix $\Dm'_{[k,\ell,q]}\in\CC^{M_{\rm sub}\times M_{\rm sub}}$ is given as
\begin{equation}
    \Dm'_{[k,\ell,q]} = \left[\Dm_{[1,1]}(\Mc_q,\Mc_q)\right]^{-1}\Dm_{[k,\ell]}(\Mc_q,\Mc_q).
\end{equation}


\begin{figure}[t]
\centering
\includegraphics[width=1.0\linewidth]{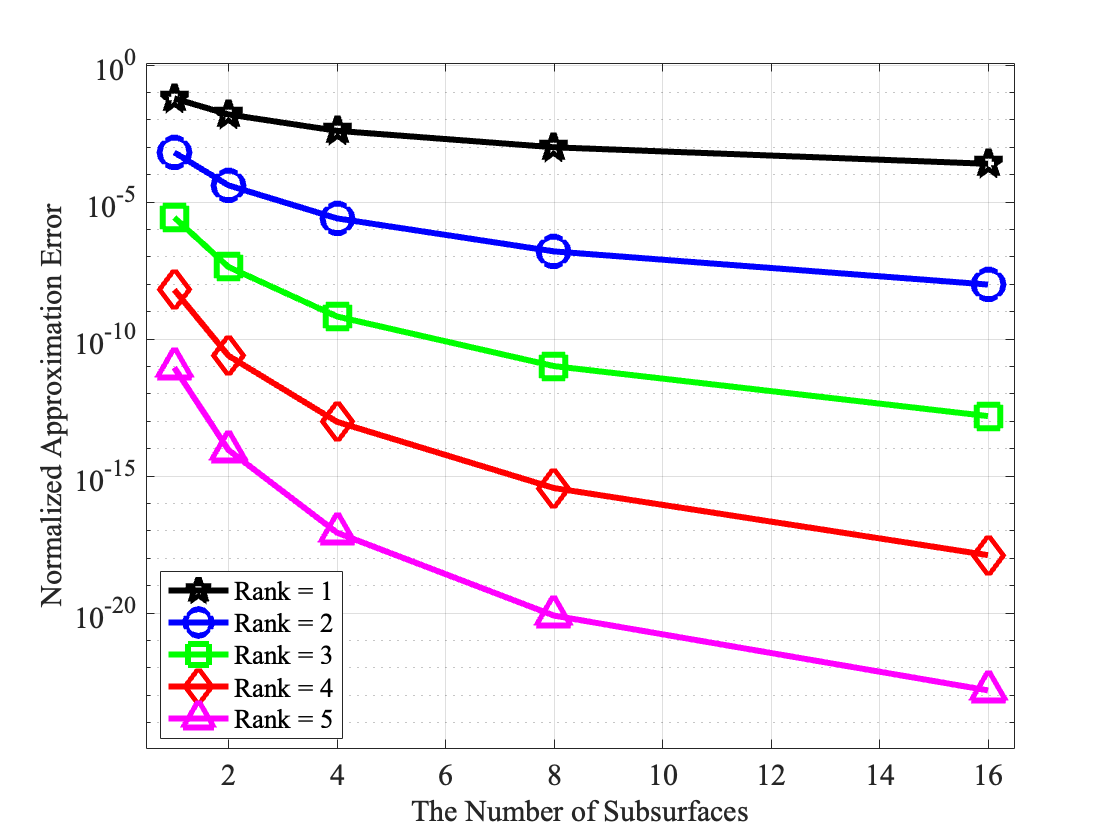}
\caption{The normalized approximation error on the number of subsurfaces, where $N=64$, $M=256$ and $L=4$.}
\end{figure}



\begin{remark}
We show that by properly choosing $Q$, the piece-wise effective channels in \eqref{eq:qeff} can be well-approximated as low-rank matrices. Let  $\tilde{\Hm}_{[k,\ell,q]}^{\rm eff}(r_q)$ denote the best rank-$r_q$ approximation of the effective channel $\Hm_{[k,\ell,q]}^{\rm eff}$. From the Eckart-Young-Mirsky theorem \cite{eckart1936approximation}, it can be easily obtained via singular value decomposition (SVD) and the normalized approximation error is computed as
\begin{equation}
   \left\|\Hm_{[k,\ell,q]}^{\rm eff}-\tilde{\Hm}_{[k,\ell,q]}^{\rm eff}(r_q)\right\|_F^2 = \sum_{i=r_q+1}^{\min(N,M_{\rm sub})}\left(\sigma_{[k,\ell,q]}(i)\right)^2.
\end{equation} where $\sigma_{[k,\ell,q]}(i)$ denotes the $i$-th largest singular value of $\Hm_{[k,\ell,q]}^{\rm eff}$. Via experiments, we measure the normalized approximation error:
\begin{equation}
      \frac{\sum_{q=1}^Q\left\|\Hm_{[k,\ell,q]}^{\rm eff}-\tilde{\Hm}_{[k,\ell,q]}^{\rm eff}(r_q)\right\|_F^2}{\left\|\Hm_{[k,\ell]}^{\rm eff}\right\|_F^2}.
\end{equation} The corresponding results are provided in Fig. 2. We observe that as the number of surfaces or the rank grows, the approximation error decreases. This reveals that the piece-wise low-rank approximation with $Q>1$ is much more accurate than the direct low-rank approximation of the effective channels (i.e., $Q=1$). In the proposed method, $Q$ is properly chosen such that $r_q$ is close to $N_{\rm RF}$, provided that the approximation error is sufficiently small. In this way, the proposed method can significantly reduce the training overhead (see Remark~\ref{remark:overhead} and Tabel 1 for details).
\hfill$\blacksquare$
\end{remark}
\vspace{0.1cm}

Harnessing the piece-wise low-rank approximation, we jointly estimate the piece-wise common column space $\{\Sm_q^{\rm RB}:q\in[Q]\}$ via collaborative low-rank approximation (CLRA). This method can guarantee a higher accuracy since $\Hm_{q}^{\rm RB}$ is well-approximated as a low-rank matrix by choosing a proper $Q$ (e.g., $Q= 4$ in our simulations). The proposed channel estimation method is named {\bf PW-CLRA}, which consists of the two phases:
\begin{itemize}
    \item {\bf Phase I.} During the first $QB$ subframes, the piece-wise column spaces $\{\Sm_{q}^{\rm RB}: q \in [Q]\}$ are estimated. The detailed methods are provided in Section~\ref{sec:ce1}.
    \item {\bf Phase II.} During the remaining $M$ subframes, the piece-wise (user- and antenna-specific) coefficient matrices $\{\Tm_{[k,\ell,q]}:q\in[Q],\; \ell\in[L]\}$ are estimated via minimum mean-square error (MMSE) estimation. The detailed methods are provided in Section~\ref{sec:ce2}.
\end{itemize} The training overhead of PW-CLRA is obtained as
\begin{equation}
    Z_{\rm Prop} = QB + M = Q(N/N_{\rm RF}) + M,\label{eq:PW-CLRA}
\end{equation} 
where the hyperparameter $Q$ can control the tradeoff between the approximation error (or the estimation accuracy) and the training overhead. 

Before describing the proposed PW-CLRA, we state the useful definition below:

\begin{definition}\label{def:2} 
Let $\Phim_{[X,Y]}$ be a $X\times Y$ orthonormal matrix satisfying the constant modulus constraints. We will construct the RF combining matrices and the reflection vectors using the matrix $\Phim_{[X,Y]}$. In practice, a DFT matrix can be adopted as the $\Phim_{[X,Y]}$. When $X=Y$, $\Phim_{[X,Y]}$ is further simplified as $\Phim_{X}$ \hfill$\blacksquare$
\end{definition}


\subsection{Piece-Wise Common Column Spaces}\label{sec:ce1}

In the first phase, we estimate the piece-wise common column spaces  $\{\Sm_{q}^{\rm RB}: q \in [Q]\}$ during the $QB$ subframes. The total cascaded channel in \eqref{eq:cascadedchannel} can be rewritten as
\begin{align}
    &\Hm_k^{\rm tot}(\vv)=\sum_{q=1}^{Q}\Hm^{\rm RB}(:,\mathcal{M}_q){\rm diag}(\vv(\mathcal{M}_q))\Hm_k^{\rm UR}(\mathcal{M}_q,:)\nonumber\\
    & \stackrel{(a)}{=}\sum_{q=1}^{Q}\Sm_q^{\rm RB}\begin{bmatrix}
    \Tm_{[k,1,q]}\vv(\mathcal{M}_q)&\cdots& \Tm_{[k,L,q]}\vv(\mathcal{M}_q)
    \end{bmatrix},\label{eq:Qsum}
\end{align} where $\vv(\Mc_q)$ denote the subvector of $\vv$ only taking the elements whose indices belong to $\Mc_q \subset [M]$ and (a) is due to the decomposition in \eqref{eq:qeff}. In the proposed method, the overall $QB$ subframes are partitioned into $Q$ groups each having $B$ subframes.

\subsubsection{Observation Processing}

In the group $q \in [Q]$, during the $B$ subframes, the BS observes:
\begin{equation}
    \{\Ym_{[z,k]}: z = B(g-1)+b,\; b\in [B]\}.
\end{equation} For ease of exposition, we re-index the observations such as
\begin{equation}
    \Ym_{[g,b, k]} = \Ym_{[B(g-1)+b, k]} \in \CC^{N_{\rm RF}\times L},\; b \in [B], g \in [Q].
\end{equation} At the subframe $b$ in the group $g$, we design the RF combining matrix as
\begin{equation}
    \Cm^{\rm 1st}_{[g,b]}= \sqrt{N}\Phim_{N}(:,\mathcal{N}_b),\label{eq:Cm}
\end{equation} where $\Nc_{b}=\{1+N_{\rm RF}(b-1),...,N_{\rm RF}b\}$. Using this construction and from \eqref{eq:variance}, we can get:
\begin{align}
    \Wm_{[g,k]}^{\rm 1st} &=  \sqrt{N}\Phim_{N}\begin{bmatrix}
        \Ym_{[g,1,k]}\\
        \vdots\\
        \Ym_{[g,B,k]}
    \end{bmatrix} = \Hm_k^{\rm tot}(\vv_g^{\rm 1st}) +\tilde{\Um}_{[g,k]}^{\rm 1st}, \label{eq:aa}
\end{align} where 
\begin{equation}
    \tilde{\Um}_{[g,k]}^{\rm 1st} \eqdef \begin{bmatrix}
        \Um_{[g,1,k]}\\
        \vdots\\
        \Um_{[g,B,k]}
    \end{bmatrix}.
\end{equation} To derive the piece-wise observations, the RIS reflection vector is constructed as
\begin{align}
    \vv_g^{\rm 1st} &=  \sqrt{Q}\begin{bmatrix}
        \Phim_{Q}(1,g){\bf 1}_{M_{\rm sub}}\\
        \vdots\\
        \Phim_{Q}(Q,g){\bf 1}_{M_{\rm sub}}
    \end{bmatrix}.\label{eq:vv}
\end{align} 
Using this construction and from \eqref{eq:Qsum} and \eqref{eq:aa}, the $\ell$-th column of $\Wm_{[g,k]}^{\rm 1st}$ can be represented as
\begin{align}
    \wv_{[g,k,\ell]}^{\rm 1st} &=  \sum_{q=1}^{Q}\Sm_q^{\rm RB}\Tm_{[k,\ell,q]}\vv_g^{\rm 1st}(\mathcal{M}_q) + \tilde{\uv}_{[g,k,\ell]}^{\rm 1st}\nonumber\\
    &= \sqrt{Q}\sum_{q=1}^{Q}\Phim_{Q}(q,g)\Sm_q^{\rm RB}\tv^{\rm sum}_{[k,\ell,q]} + \tilde{\uv}_{[g,k,\ell]}^{\rm 1st},
\end{align} where $\tilde{\uv}_{[q,k,\ell]}^{\rm 1st}$ is the $\ell$-th column of $\tilde{\Um}_{[g,k]}^{\rm 1st}$ and
\begin{equation}
    \tv_{[k,\ell,q]}^{\rm sum}\eqdef\Tm_{[k,\ell,q]}{\bf 1}_{M_{\rm sub}}.
\end{equation}
Note that from \eqref{eq:Tklq}, we can get
\begin{align}
    \tv_{[k,\ell,q]}^{\rm sum} &= \Tm_{[k,\ell,q]}{\bf 1}_{M_{\rm sub}}\nonumber\\
    &=\Tm_{q}^{\rm RB}\mbox{diag}(\Hm_{k}^{\rm UR}(:,\ell))(\Mc_q,\Mc_q){\bf 1}_{M_{\rm sub}}\nonumber\\
    &=\Tm_{q}^{\rm RB}\Hm_{k}^{\rm UR}(\Mc_q, \ell).\label{eq:tsum}
\end{align} Concatenating $\{\wv_{[g,k,\ell]}^{\rm 1st}: g \in [Q]\}$, we can obtain:
\begin{align}
    {\Vm}_{[\ell,k]}^{\rm 1st}&= \frac{1}{\sqrt{Q}}\begin{bmatrix}
        \wv_{[1,k,\ell]}^{\rm 1st} &\cdots& \wv_{[Q,k,\ell]}^{\rm 1st}
    \end{bmatrix}\Phim_{Q}^{\herm}\nonumber\\
    &=\left[\Sm_1^{\rm RB}\tv_{[k,\ell,1]}^{\rm sum}\;\cdots\;\Sm_Q^{\rm RB}\tv_{[k,\ell,Q]}^{\rm sum}\right]\nonumber\\
&\quad\quad\quad\quad\quad\quad+ \frac{1}{\sqrt{Q}}\left[\tilde{\uv}_{[1,k,\ell]}^{\rm 1st}\;\cdots\;\tilde{\uv}_{[Q,k,\ell]}^{\rm 1st}\right]\Phim_{Q}^{\herm}.\label{eq:Qbeam}
\end{align} Remarkably, the noise power is reduced as the number of subsurfaces increases attaining a beamforming gain from the RIS. Taking every $q$-th column of  $\{{\Vm}_{[\ell,k]}^{\rm 1st}:\ell\in[L]\}$, we can derive the observations for estimating $\Sm_{q}^{\rm RB}$ as follows:
\begin{align}
    \Mm_{[q,k]}^{\rm 1st} &\eqdef\Sm_q^{\rm RB}\begin{bmatrix}
        \tv_{[k,1,q]}^{\rm sum} &\cdots& \tv_{[k,L,q]}^{\rm sum}
    \end{bmatrix} + {\Nm}_{[q,k]}^{\rm 1st}\nonumber\\
    &\stackrel{(a)}{=}\Sm_{q}^{\rm RB}\Tm_{q}^{\rm RB}\Hm_{k}^{\rm UR}(\Mc_q,:)+ {\Nm}_{[q,k]}^{\rm 1st}\nonumber\\
    &=\Hm_{q}^{\rm RB}\Hm_{k}^{\rm UR}(\Mc_q,:)+ {\Nm}_{[q,k]}^{\rm 1st},\; k \in [K],
\end{align}  where (a) follows from \eqref{eq:tsum} and
\begin{equation}
    {\Nm}_{[q,k]}^{\rm 1st}\eqdef\left[\tilde{\uv}_{[q,k,1]}^{\rm 1st}\;\cdots\;\tilde{\uv}_{[q,k,L]}^{\rm 1st}\right]\Phim_{Q}^{\herm}.
\end{equation} Since the $K$ users share the piece-wise common column space (i.e., $\Sm_{q}^{\rm RB}$), we can jointly optimize it from the observations $\{\Mm_{[q,k]}^{\rm 1st}: k \in [K]\}$. These observations can be expressed as follows:
\begin{align}
    \Mm_q^{\rm 1st} &\eqdef \begin{bmatrix}
        \Mm_{[q,1]}^{\rm 1st} &\cdots& \Mm_{[q,K]}^{\rm 1st}
    \end{bmatrix}\nonumber\\
    &=\Hm_{q}^{\rm RB}\begin{bmatrix}
        \Hm_{1}^{\rm UR}(\Mc_q,:) &\cdots& \Hm_{K}^{\rm UR}(\Mc_q,:)
    \end{bmatrix}+ \Nm_q^{\rm 1st}\nonumber\\
    &=\Hm_{q}^{\rm RB}\Hm^{\rm UR}(\Mc_q,:)+\Nm_q^{\rm 1st},\label{eq:1st}
\end{align} where 
\begin{align}
    \Hm^{\rm UR}&\eqdef \left[\Hm_{1}^{\rm UR}\;\cdots\;\Hm_{K}^{\rm UR}\right]\in\CC^{M\times KL}\\
    \Nm_q^{\rm 1st}&\eqdef\left[{\Nm}_{[q,1]}^{\rm 1st}\;\cdots\;{\Nm}_{[q,K]}^{\rm 1st}\right]\in\CC^{N\times KL}.
\end{align} From the fundamental theory of linear algebra \cite{horn2012matrix}, we can see that the column space of $\Mm_{q}^{\rm 1st}$ is contained in the column space of $\Hm_{q}^{\rm RB}$, provided that the impact of the additive noise is very small. Also, if $\mbox{rk}(\Hm^{\rm UR}(\Mc_q,:))=M_{\rm sub}$, then they can be equivalent. To estimate the column space of $\Hm_{q}^{\rm RB}$ exactly,  we have the necessary condition
\begin{equation}
    M_{\rm sub} \leq KL.
\end{equation} Equivalently, given a system configuration $(M,K,L)$, the hyperparameter $Q$ should be chosen such that
\begin{equation}
    Q \geq \frac{M}{KL}.\label{eq:constraint_Q}
\end{equation} Even if $\mbox{rk}(\Hm^{\rm UR}(\Mc_q,:))<M_{\rm sub}$ (i.e., $Q<M/KL$), the column space of $\Mm_{q}^{\rm 1st}$ would be a good estimate of the column space of $\Hm_{q}^{\rm RB}$ (i.e., $\Sm_{q}^{\rm RB}$). In practice, to reduce the training overhead while achieving a target estimation accuracy, we commonly choose the $Q$ which is smaller than $M/KL$. Obviously, the estimation accuracy can improve as $\mbox{rk}(\Hm^{\rm UR}(\Mc_q,:))$ approaches to $M_{\rm sub}$ and the impact of the additive noise is negligible. Remarkably, as the number of users (i.e., $K$) increases, $\mbox{rk}(\Hm^{\rm UR}(\Mc_q,:))$ is close to $M_{\rm sub}$ with high probability. This comes from the fact that the channel responses among users are statistically independent with high probability since each user experiences different channel characteristics (i.e., the distance or angle of arrival/departure). Thus, the growth of the users brings a performance gain, which is referred to as the multi-user gain throughout the paper.


\subsubsection{Estimation}
We explain how to estimate the piece-wise common column spaces (i.e., $\{\Sm_q^{\rm RB}:q\in[Q]\}$) from the processed observations $\{\Mm_{q}^{\rm 1st}:q\in[Q]\}$ in \eqref{eq:1st}. Note that the proposed estimation performs in parallel with respect to the subsurface $q \in [Q]$. 
From $\Mm_{q}^{\rm 1st}$, we can estimate $\Sm_{q}^{\rm}$ following the subspace estimation method proposed in \cite{Lee2024near} (called collaborative low-rank approximation (CLRA)), which is composed of the two steps:
\begin{enumerate}
    \item From the processed observation $\Mm_q^{\rm 1st}$, the rank of $\Sm_q^{\rm RB}$ is estimated via the minimum description length (MDL) criterion (see \cite{Lee2024near} for details). The estimated rank is referred to as 
    \begin{equation}
        \hat{r}_q=\hat{\mbox{rk}}(\Sm_{q}^{\rm RB}).
    \end{equation}
    \item From the eigenvalue decomposition of $\Mm_q^{\rm 1st}\left(\Mm_q^{\rm 1st}\right)^{\herm} = \tilde{\Sm}_q\Sigmam_q\tilde{\Sm}_q^{\herm}$, the estimated $\Sm_{q}^{\rm RB}$ is simply obtained as
    \begin{equation}\label{eq:estimated_S}
    \hat{\Sm}_q^{\rm RB}=\tilde{\Sm}_q(:,[\hat{r}_q])\in\CC^{N\times \hat{r}_q}.
\end{equation}
\end{enumerate}

\subsection{Piece-Wise Coefficient Matrices}\label{sec:ce2}

In the second phase, we estimate the piece-wise (user- and antenna-specific) coefficient matrices $\{\Tm_{[k,\ell,q]}:q\in[Q], \ell\in[L]\}$. 

\subsubsection{Observation Processing} During the remaining $M$ subframes, the BS receives
\begin{equation}
    \{\Ym_{[m_0+m,k]}: m \in [M]\},
\end{equation} where $m_0=QB$ denotes the index of the last subframe in the first part. To ease of exposition, we let
\begin{equation}
    \Wm_{[m,k]}^{\rm 2nd} = \Ym_{[m_0+m,k]},\; m \in [M].
\end{equation} For each subframe $m \in [M]$, the RF combining matrix and the RIS reflection vector are respectively constructed as
\begin{align}
    &\Cm^{\rm 2nd}_{m} =  \sqrt{N}\Phi_{[N,N_{\rm RF}]},\nonumber\\
    &\vv_{m}^{\rm 2nd} =  \sqrt{M}\Phi_{M}(:,m).
\end{align} Leveraging these constructions, we can get:
\begin{align}
    &\Vm_{[\ell,k]}^{\rm 2nd}\eqdef \frac{1}{\sqrt{NM}}\begin{bmatrix}
        \wv_{[1,k,\ell]}^{\rm 2nd} &\cdots& \wv_{[M,k,\ell]}^{\rm 2nd}
    \end{bmatrix}\Phim_{M}^{\herm}\nonumber\\
    &=\Phim_{[N,N_{\rm RF}]}^{\herm}\Hm_{[k,\ell]}^{\rm eff}+\Nm_{[\ell,k]}^{\rm 2nd},\label{eq:2ndobs}
\end{align} where $\wv_{[1,k,\ell]}^{\rm 2nd}$ denotes the $\ell$-th column of $\Wm_{[m,k]}^{\rm 2nd}$ and 
\begin{equation}
    \Nm_{[\ell,k]}^{\rm 2nd}\eqdef \frac{1}{\sqrt{M}}\Phim_{[N,N_{\rm RF}]}^{\herm}\left[\tilde{\uv}_{[1,k,\ell]}^{\rm 2nd}\;\cdots\; \tilde{\uv}_{[Q,k,\ell]}^{\rm 2nd}\right]\Phim_{M}^{\herm}.\label{eq:np}
\end{equation} Note that in \eqref{eq:2ndobs}, the noise power is reduced as the number of RIS elements (i.e., $M$) increases.

Finally, we can derive the observations to estimate the piece-wise coefficient matrices:
\begin{align}
    \Mm_{[\ell,q,k]}^{\rm 2nd}&\eqdef\Vm_{[\ell,k]}^{\rm 2nd}(:,\mathcal{M}_q)\nonumber\\
    &=\Phim_{[N, N_{\rm RF}]}^{\herm}\Hm_{[k,\ell,q]}^{\rm eff} + \Nm_{[\ell,k]}^{\rm 2nd}(:,\mathcal{M}_q)\nonumber\\
    &=\Phim_{[N, N_{\rm RF}]}^{\herm}
        \Sm_q^{\rm RB}\Tm_{[k,\ell,q]} + \Nm_{[\ell,k]}^{\rm 2nd}(:,\mathcal{M}_q).\label{eq:2nd}
\end{align}

\subsubsection{Estimation} From \eqref{eq:2nd}, we can easily estimate $\Tm_{[k,\ell,q]}$ via least square (LS) estimation, which is given by
\begin{align}
    \hat{\Tm}_{[k,\ell,q]}^{\rm LS} &\eqdef {\Pm}_q^{\dag}\Mm_{[\ell,q,k]}^{\rm 2nd}\in\CC^{\hat{r}_q\times M_{\rm sub}},\label{eq:TLS}
\end{align} where ${\Pm}_{q} \eqdef \Phim_{[N\times N_{\rm RF}]}^{\herm}\hat{\Sm}_q^{\rm RB}\in\CC^{N_{\rm RF}\times \hat{r}_q}$ is the projection matrix and note that $\hat{\Sm}_q^{\rm RB}$ is estimated in the first part. For the existence of the pseudo-inverse $\Pm_q^{\dag}$, the following condition should hold:
\begin{equation}
    \hat{r}_q \leq N_{\rm RF},\; \forall q \in [Q],
\end{equation} This condition is satisfied if $Q$ is properly chosen such that $Q\geq M/N_{\rm RF}$. We can further improve the estimation accuracy at lower SNR regimes via minimum mean-square error (MMSE) estimation. Note that to the multiplication of unitary matrices, the elements of $\Nm_{[\ell,k]}^{\rm 2nd}(:,\mathcal{M}_q)$ follow independently identically circularly symmetric complex Gaussian distribution with mean zero and variance $\sigma^2$. Taking the impact of noise into account, from \cite{kailath2000linear}, the MMSE solution is obtained as
\begin{equation}
    \Tm_{[k,\ell,q]}^{\rm MMSE}= \left(\Pm_q^{\herm}\Pm_q + \frac{KL}{\rho_k}\Id_{\hat{r}_q}\right)^{-1}\Mm_{[\ell,q,k]}^{\rm 2nd},\label{eq:MMSEsol}
\end{equation} where $\rho_k=P\left\|\Hm_{k}^{\rm tot}(\vv)\right\|_F^2/\sigma^2$. 
From \eqref{eq:estimated_S} and \eqref{eq:MMSEsol}, we can obtain the estimated column-wise effective channel in \eqref{eq:CW} such as
\begin{equation}
    \hat{\Hm}_{[k,\ell]}^{\rm eff} = \begin{bmatrix}
        \hat{{\Hm}}_{[k,\ell,1]}^{\rm eff} &\cdots& \hat{{\Hm}}_{[k,\ell,Q]}^{\rm eff}
    \end{bmatrix},
\end{equation} with $\hat{{\Hm}}_{[k,\ell,q]}^{\rm eff} = \hat{\Sm}_q^{\rm RB}\hat{\Tm}_{[k,\ell,q]}^{\rm MMSE}$. 

\vspace{0.1cm}
In the near-field channel estimation for the XL-RIS assisted XL-MIMO system, a high-SNR assumption is reasonable due to the extremely large array gain and the shorter distance between a transmitter and a receiver \cite{Ayach2014}. Focusing on high-SNR ranges, we can improve the estimation accuracy on the piece-wise coefficient matrices using the scaling property in \eqref{eq:scaling}. Namely, they should satisfy the following equalities:
\begin{equation}
    \Tm_{[k,\ell,q]} = \Tm_{[1,1,q]}\Dm'_{[k,\ell,q]}
\end{equation} for $\ell \in [L]$ and $k \in [K]$. By taking this structure into account, we can refine the MMSE estimations in \eqref{eq:MMSEsol}. The corresponding joint optimization (JO) is formulated as follows:
\begin{equation}
\min_{\hat{\Am}_q,\hat{\mathcal{D}}'_{q}}\sum_{k=1}^{K}\sum_{\ell=1}^{L}\left\|\hat{\Tm}_{[k,\ell,q]}^{\rm MMSE}-\hat{\Am}_q\hat{\Dm}'_{[k,\ell,q]}\right\|_F^2,
\end{equation} where $\hat{\Am}_q=\hat{\Tm}_{[1,1,q]}^{\rm MMSE}$ and
\begin{equation}
    \hat{\mathcal{D}}'_{q} = \{\hat{\Dm}'_{[k,\ell,q]}:k\in[K],\ell\in[L]\}.
\end{equation} As derived in \cite{Lee2024near}, the above optimization can be solved via an iterative algorithm. At each iteration  $t\in[t_{\rm max}]$, where $t_{\rm max}$ denotes the number of maximum iterations), the closed-form solutions are respective obtained as
\begin{align}
    &\hat{\Dm}'^{(t)}_{[k,\ell,q]} = \diag\left(d_{[k,\ell,q,1]}^{(t)},\dots,d_{[k,\ell,q,M_{\rm sub}]}^{(t)}\right),\label{eq:Dproj}\\
    &\hat{\Am}_q^{(t)} = \left(\sum_{k=1}^{K}\sum_{\ell=1}^{L}\hat{\Tm}_{[k,\ell,q]}^{\rm MMSE}\hat{\Dm}'^{(t)}_{[k,\ell,q]}\right)\nonumber\\
    &\quad\quad\quad\quad\quad\quad\times\left(\sum_{k=1}^{K}\sum_{\ell=1}^{L}\left(\hat{\Dm}'^{(t)}_{[k,\ell,q]}\right)^{\herm}\hat{\Dm}'^{(t)}_{[k,\ell,q]}\right)^{-1}\label{eq:Aleast},
\end{align} with initial value $\hat{\Am}_q^{(0)}=\hat{\Am}_q$ and 
\begin{equation}
    d_{[k,\ell,q,m]}^{(t)}=\left(\hat{\Am}_q^{(t-1)}(:,m)\right)^{\dag}\hat{\Tm}_{[k,\ell,q]}^{\rm MMSE}(:,m).
\end{equation} From the joint optimization, the estimated column-wise effective channels are obtained as follows:
\begin{align}
    \hat{\Hm}_{[k,\ell,q]}^{\rm eff} &= \hat{\Sm}_q^{\rm RB}\hat{\Tm}_{[k,\ell,q]}^{\rm JO}=\hat{\Sm}_q^{\rm RB}\hat{\Am}_q^{(t_{\rm max})}\hat{\Dm}'^{(t_{\rm max})}_{[k,\ell,q]}.
\end{align} The proposed CLRA-JO using the joint optimization is referred to as {\bf PW-CLRA-JO}.

\begin{algorithm}[h]
\caption{The Proposed PW-CLRA-JO}
\begin{algorithmic}[1]

\State {\bf Input:} $\{\hat{\Sm}_{q}^{\rm RB}:q\in[Q]\}$ in \eqref{eq:estimated_S}, $\{\Mm_{[\ell,q,k]}^{\rm 2nd}: \ell\in[L],q\in[Q],k\in[K]\}$ in \eqref{eq:2nd}, and the maximum number of iterations $t_{\rm max}$.
\vspace{0.1cm}

\vspace{0.1cm}
\State {\bf Initialization:} Set $\hat{\Am}_q^{(0)} = \hat{\Tm}^{\rm MMSE}_{[1,1,q]}$ from \eqref{eq:MMSEsol} and $t=0$.
\vspace{0.1cm}
\State {\bf Repeat until $t=t_{\rm max}$}
\begin{itemize}
    \item $t=t+1$.
    \item Given $\hat{\Am}_q^{(t-1)}$, update the $\hat{\Dm}'^{(t)}_{[k,\ell,q]}$ via \eqref{eq:Dproj}.
    \item Given $\hat{\Dm}'^{(t)}_{[k,\ell,q]}$, update $\hat{\Am}_q^{(t)}$  via \eqref{eq:Aleast}.
\end{itemize}
\vspace{0.1cm}
\State {\bf Output:} The estimated effective channels:
\begin{equation*}
    \hat{\Hm}_{[k,\ell,q]}^{\rm eff} = \hat{\Sm}_{q}^{\rm RB}\hat{\Am}_q^{(t_{\rm max})}\hat{\Dm}'^{(t_{\rm max})}_{[k,\ell,q]}.
\end{equation*}
\end{algorithmic}
\end{algorithm}

\vspace{0.2cm}
\begin{remark}\label{remark:overhead}
We consider the RIS-assisted MU-MIMO system, where $M=256$, $N=128$, $N_{\rm RF}=16$, $K=8$, and $L=2$. Suppose that $\hat{\mbox{rk}}(\Hm^{\rm RB})=64$. This implies that the column space of $\Hm^{\rm RB}$ can be well-approximated as $\hat{\mbox{rk}}(\Hm^{\rm RB})$-dimensional subspace. From \eqref{eq:CLRA_JO}, the CLRA-JO in \cite{Lee2024near} requires the training overhead such as 
    \begin{align*}
        Z_{\rm CLRA}&=16\times (128/16)+4\times 256 = 1104,
    \end{align*} where we chose $B_c=16$ and $B_r=4$ as the minimum values to meet the constraints in \eqref{CLRA_JO:constraint}. We remark that these values are larger than those in \cite{Lee2024near}  because our near-field LoS channel has higher rank than the far-field counterpart in \cite{Lee2024near}. From \eqref{eq:PW-CLRA}, the proposed method with $Q=16$ requires the training overhead as
    \begin{align*}
        Z_{\rm Prop} = 16 \times (128/16) + 256 = 336,
    \end{align*} where we chose $Q=16$ as the minimum value to satisfy the constraint in \eqref{eq:constraint_Q}. In this comparison, we can see that the proposed method can significantly reduce the training overhead compared with the SOTA method by $71\%$.\flushright$\blacksquare$
\end{remark}

\subsection{Complexity Analysis}\label{sec:ComplexityAnalysis}

We analyze the computational complexity of the proposed PW-CLRA-JO in Algorithm 1. Following the related works in  \cite{chen2023channel, Lee2024near}, the computational complexity is measured by counting the number of complex multiplication (CM). Regarding the complexity of the piece-wise subspace estimation in Section~\ref{sec:ce1}, we need to compute the eigenvalue decomposition in \eqref{eq:estimated_S}, which requires the complexity of $\mathcal{O}(Q N^3)$ for each $q \in [Q]$. Also, it is required to compute the matrix inversion in \eqref{eq:MMSEsol} with the complexity of $\mathcal{O}(\hat{r}_q^3)$ for each $q \in [Q]$. Lastly, in the second part of the training, we need to solve the joint optimizations in \eqref{eq:Dproj} and \eqref{eq:Aleast}, which respectively require the complexities of $\Delta_{D}$ and $\Delta_{A}$ where
\begin{align}
    \Delta_{D}&\eqdef KLM_{\rm sub}\sum_{q=1}^{Q}(3\hat{r}_q+1),\label{eq:D}\\
    \Delta_{A}&\eqdef KLM_{\rm sub}\sum_{q=1}^{Q}(\hat{r}_q+2).\label{eq:T}
\end{align} The overall computational complexity of the PW-CLRA-JO is obtained as
\begin{align}
    \Psi_{\rm Prop} = \mathcal{O}\left(QN^3 + \sum_{q=1}^{Q}\hat{r}_{q}^3 + t_{\rm max}(\Delta_{D} + \Delta_{A})\right).
\end{align}

\begin{figure}[t]
\centering
\includegraphics[width=0.9\linewidth]{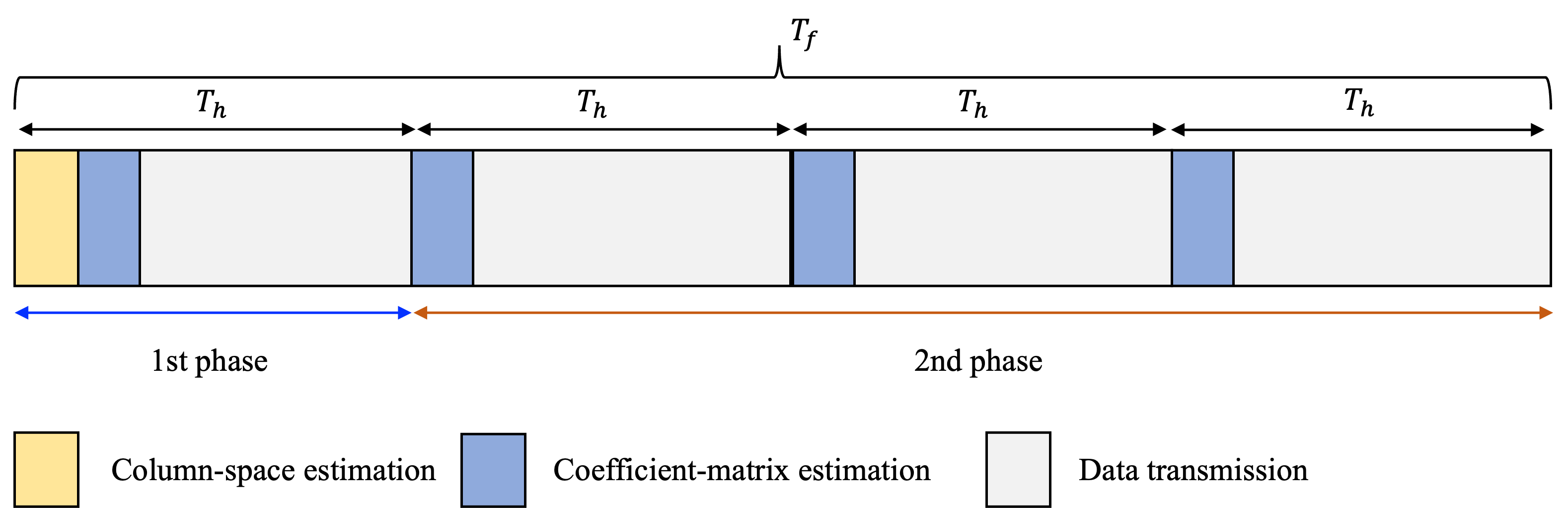}
\caption{Two-phase communication protocol.}
\label{fig:two-phase}
\end{figure}

Because of the piece-wise estimation, in the first phase of the training, the proposed PW-CLRA-JO has the $Q$ times higher computational complexity than the CLRA-JO in \cite{Lee2024near}, namely, the complexities of CLRA-JO and PW-CLRA-JO are equal to $\Oc(N^3)$ and $\Oc(Q N^3)$, respectively. On the other hand, they have almost the same complexity in the second phase of the training. We emphasize that since $Q$ is chosen between 2 and 4 in many practical scenarios, the additional complexity to handle the near-field LoS component is affordable. As highlighted in \cite{Lee2024near}, in practical systems, the coherence time of the BS-RIS channel (denoted by $T_f$) is commonly much longer than that of the RIS-User channels (denoted by $T_h$). Leveraging this fact, we can further reduce the computational complexity and the training overhead of the PW-CLRA-JO via the two-phase communication protocol in Fig.~\ref{fig:two-phase}. Under this protocol, the impact of the first-part complexity of $\mathcal{O}\left(Q N^3\right)$ (or  $\mathcal{O}\left(N^3\right)$) can be negligible when $T_f \gg T_h$. Consequently, the overall complexities of the PW-CLRA-JO and CLRA-JO become almost the same. Furthermore, the average training overhead of the PW-CLRA-JO in \eqref{eq:PW-CLRA} can be reduced to $M$ instead of $Q(N/N_{\rm RF})+M$.

\begin{figure*}[!t]
    \centering
    \subfigure[$K=1$]{
        \includegraphics[width=0.48\linewidth]{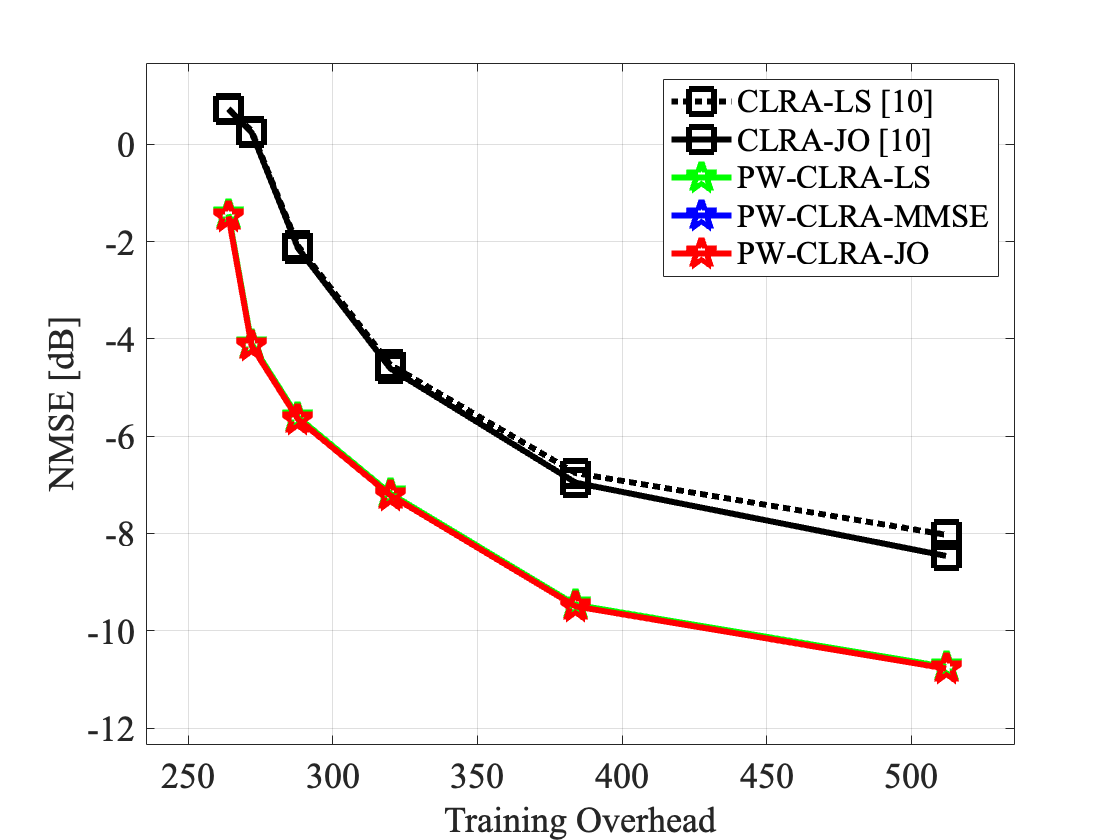}
    }
    \subfigure[$K=4$]{
        \includegraphics[width=0.48\linewidth]{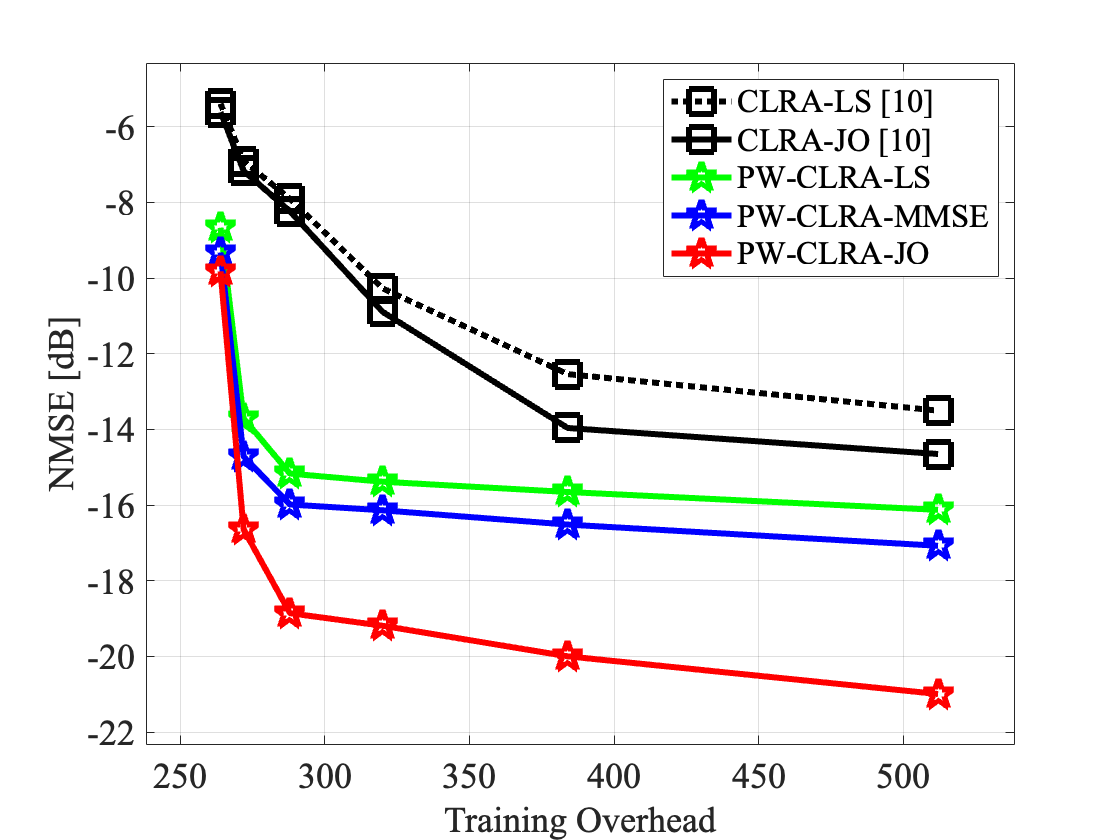}
    }
    \caption{The NMSE on the training overhead. ${\rm SNR}=10\;{\rm dB}$ and $d_{x}^{\rm RB}=50\;{\rm m}$.} \label{fig:action_tabular}
\end{figure*}

\section{Simulation Results}\label{sec:SR}

We conduct the simulations to verify the effectiveness of the proposed PW-CLRA for RIS-assisted MU-MIMO systems, where $N=128$, $N_{\rm RF} = 16$, $N_{\rm NLoS}^{\rm RB} = 3$, $M=256$, and $L=2$. Following a 3-D Cartesian coordinate system, the BS and the RIS are located at $(d_x^{\rm RB}, 20, 5)\;{\rm m}$ and $(0, 0, 10)\;{\rm m}$, respectively. For simplicity, we determine the distance between the BS and the RIS only controlling the $x$-coordinate value $d_x^{\rm RB}$. In order to generate various channel environments, each user $k \in [K]$ is located at $(d_{[x,k]}^{\rm UR},-20,5)\;{\rm m}$, where $d_{[x,k]}^{\rm UR}$ is uniformly and randomly selected from $[20,30]{\rm m}$. In our simulations, the the system carrier frequency is set by $f_c=50\;{\rm GHz}$ with $\lambda=0.006\;{\rm m}$ and following the simulation setting in \cite{yang2023}, the noise power is set by $\sigma^2=-169\;{\rm dBm}$.
As in the related works  \cite{chen2023channel,He2021, yang2023}, we adopt the normalized mean square error (NMSE) to evaluate a channel estimation accuracy:
\begin{equation}
    \mbox{NMSE} \eqdef \EE\left[\frac{1}{K} \sum_{k=1}^{K} \frac{\|\hat{\Hm}_{k}^{\rm eff} - \Hm_{k}^{\rm eff}\|_F^2}{\|\Hm_{k}^{\rm eff}\|_F^2} \right],
\end{equation} where the expectation is evaluated by Monte Carlo simulations with $10^3$ trials. To verify the effectiveness of the proposed {\bf PW-CLRA-JO}, we consider the following benchmark methods:
\begin{itemize}
    \item {\bf CLRA-JO} \cite{Lee2024near}: This can be regarded as the SOTA method. Also,  the {\bf CLRA-LS} is considered as the low-complexity version of the CLRA-JO.
    \item {\bf 2D-LS} \cite{Yang2024}: In this method, assuming that the RIS-BS channel is perfectly known, the RIS-User channels are estimated via least-square (LS) estimation. This method is not practical but can provide the lower-bound of the estimation accuracy (or the performance upper-bound).
    \item {\bf PW-CLRA-LS}: In the proposed PW-CLRA-JO, the piece-wise coefficient matrices are simply derived via LS estimation in  \eqref{eq:TLS}. 
    \item {\bf PW-CLRA-MMSE}: In the proposed PW-CLRA-JO, the piece-wise coefficient matrices are obtained by MMSE estimation in \eqref{eq:MMSEsol}.
\end{itemize} Considering practical scenarios, the training overhead is assumed to be less than or equal to $512$ (i.e., $Z \leq 512$). Accordingly, the hyperparameter of the proposed PW-CLRA is chosen as $Q \in \{1,2,4,8,16,32\}$. Regarding the hyperparameters of the CLRA (denoted by $B_c$ and $B_r$ in \eqref{eq:CLRA_JO}), $B_r=1$ is the only choice for  $Z \leq 512$ and $B_c$ is chosen as $Q$ to ensure the same training overhead with the PW-CLRA.

\begin{table}[ht]
\caption{The required training overhead for achieving a target accuracy. $K=4$, ${\rm SNR}=10\;{\rm dB}$ and $d_{x}^{\rm RB}=50\;{\rm m}$.}
\setlength{\tabcolsep}{5pt}
\renewcommand{\arraystretch}{1.5}

 \centering

 \begin{tabular}
  {P{70pt}|P{30pt}|P{50pt}|P{60pt}}
  \toprule[1.5pt]
   \hline
  Target Accuracy & -12 dB & -16 dB & -20 dB\\ 
\hline
 CLRA-LS [10] & 370 & 782 $(B_r>1)$ & 1,294 $(B_r>1)$ \\ 

 {\bf CLRA-JO [10]} & 340 & 752 $(B_r>1)$ & {\bf 1,264 $(B_r>1)$} \\

  PW-CLRA-LS & 260 & 470 & Saturated \\
  PW-CLRA-MMSE & 260 & 288 & Saturated \\

  {\bf PW-CLRA-JO} & 260 & 270 & {\bf 380} \\
\hline
 \bottomrule[1.5pt]
\end{tabular}
\end{table}

Fig. 4 shows the NMSE as a function of the training overhead $Z$. We observe that the proposed PW-CLRA methods significantly outperform the SOTA methods (i.e., the CLRA methods). This indeed demonstrates the effectiveness of the piece-wise collaborative low-rank approximation for high-rank near-field LoS/NLoS channel estimations. Comparing the performances of the multi-user case in Fig. 4 (b) with the single-user case in Fig. 4 (a), we identify that the proposed method can attain a multi-user gain. For example, in the multi-user case (i.e., $K=4$), the PW-CLRA-JO requires about $270$ subframes to achieve the target accuracy ${\rm NMSE}=-17\;{\rm dB}$. Whereas, in the single-user case, this method achieves the ${\rm NMSE}=-4\;{\rm dB}$ having the same training overhead. Due to the multi-user gain, the estimation accuracy can be rapidly enhanced only using a very small number of training overhead. Table I also validates the superiority of the proposed method compared with the SOTA methods. We can see that given a target estimation accuracy, the proposed PW-CLRA-JO requires much smaller training overhead than the SOTA method (i.e., the CLRA-JO in \cite{Lee2024near}). When the target accuracy is ${\rm NMSE} = -20\;{\rm dB}$, the proposed method can reduce the training overhead about $70\%$ compared with the SOTA method. This is well-matched with our analysis in Remark 3.

\begin{figure}[t]
\centering
\includegraphics[width=1.0\linewidth]{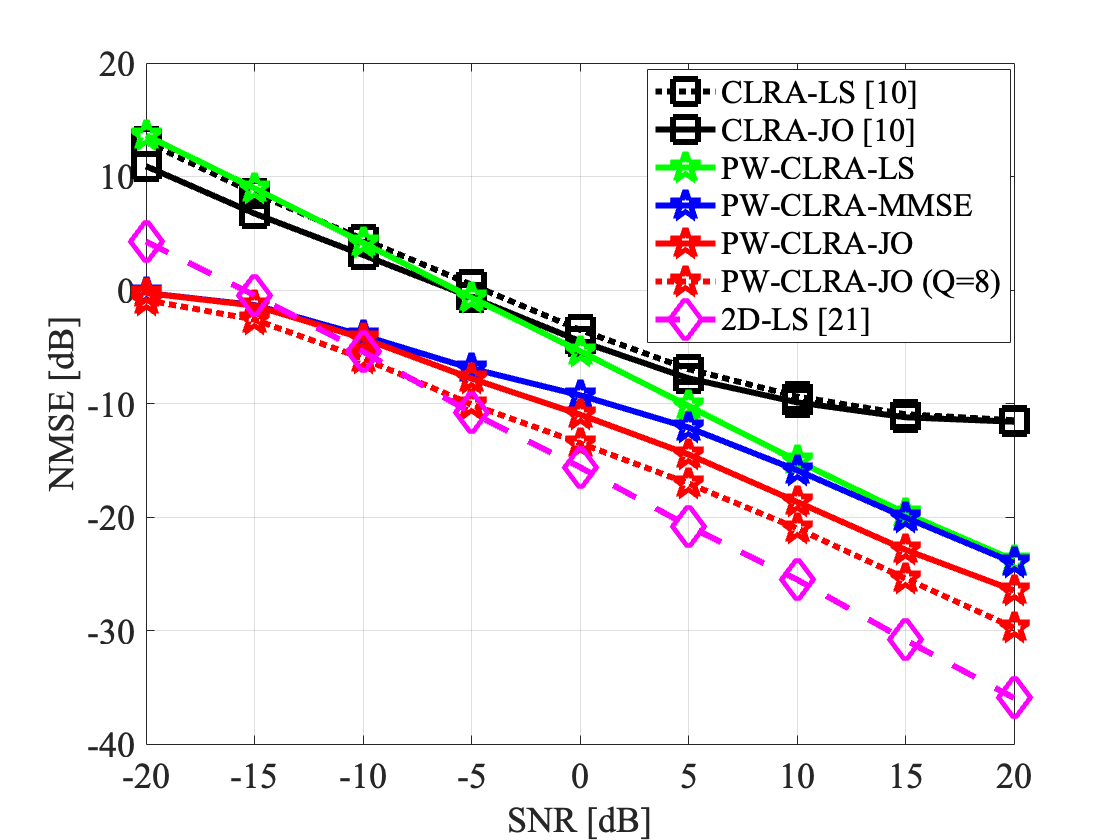}
\caption{The NMSE on the SNR. $K=4$ and $d_{x}^{\rm RB}=50\;{\rm m}$.}
\end{figure}

\begin{figure}[t]
\centering
\includegraphics[width=1.0\linewidth]{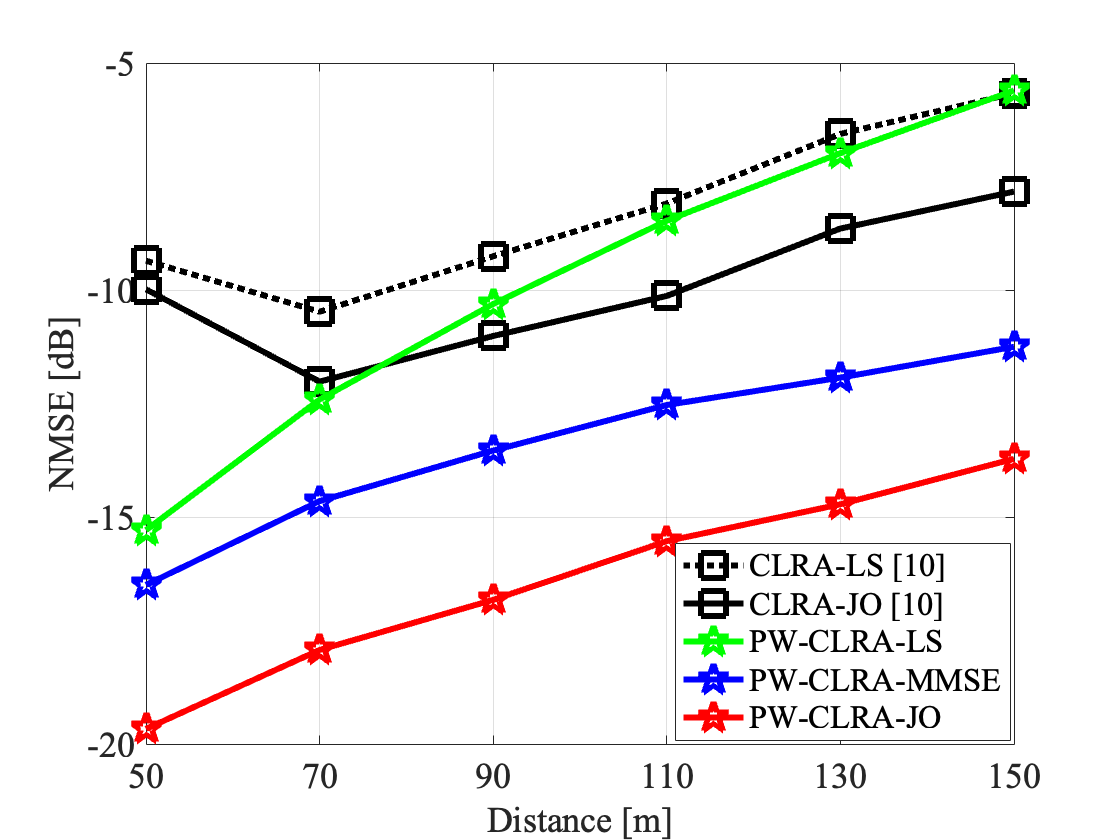}
\caption{The NMSE on the distance $d_{x}^{\rm RB}$, $K=B_c=Q=4$ and $B_r = 1$.}
\end{figure}

Fig. 5 shows the NMSE on the SNR. As expected, in low-SNR regimes (i.e., $\SNR=-20\sim 5\;{\rm dB}$), the PW-CLRA-MMSE has a significant gain compared with the PW-CLRA-LS as the MMSE estimation can well address the noise-boosting problem. The proposed PW-CLRA-JO can outperform the benchmark methods, which verifies that the joint optimization based on the scaling property is indeed beneficial. Noticeably, the estimation accuracy of the PW-CLRA methods is rapidly enhanced as SNR increases, whereas the CLRA methods suffer from an error-floor problem. Namely, the proposed method can provide stable performances over a wide range of SNRs. Although the proposed PW-CLRA-JO has a performance gap from the 2D-LS in high-SNRs, this gap can be reduced as $Q$ increases. We emphasize that the 2D-LS is not practical at all because it requires a prior knowledge on the BS-RIS channel (i.e., $\Hm^{\rm RB}$). Hence, this method can only be considered as the performance-limit. Compared with this bound, we can see that the proposed PW-CLRA-JO yields an attractive performance while having lower computational complexity and training overhead. Furthermore, they can be significantly reduced via two-phase communication protocol in Section~\ref{sec:ComplexityAnalysis}.

Fig. 6 shows the NMSE as a function of the BS-RIS distance, in which for all distances, the transmit power is fixed such that $\SNR=10\;{\rm dB}$ at $d_{x}^{\rm RB} = 50\;{\rm m}$. It is clear that the rank of the near-field LoS channel increases as the BS-RIS distance gets closer, because the distance discrepancy within a subframe becomes larger. We observe that as the distance decreases, the estimation accuracy of the proposed method is enhanced due to the power-gain (i.e., the increased SNR). This verifies that the proposed method indeed performs well under high-rank BS-RIS channels. In contrast, the SOTA method is not suitable for the high-rank LoS channel since its performance degradation is notable in the shorter distances.

\begin{figure}[t]
\centering
\includegraphics[width=1.0\linewidth]{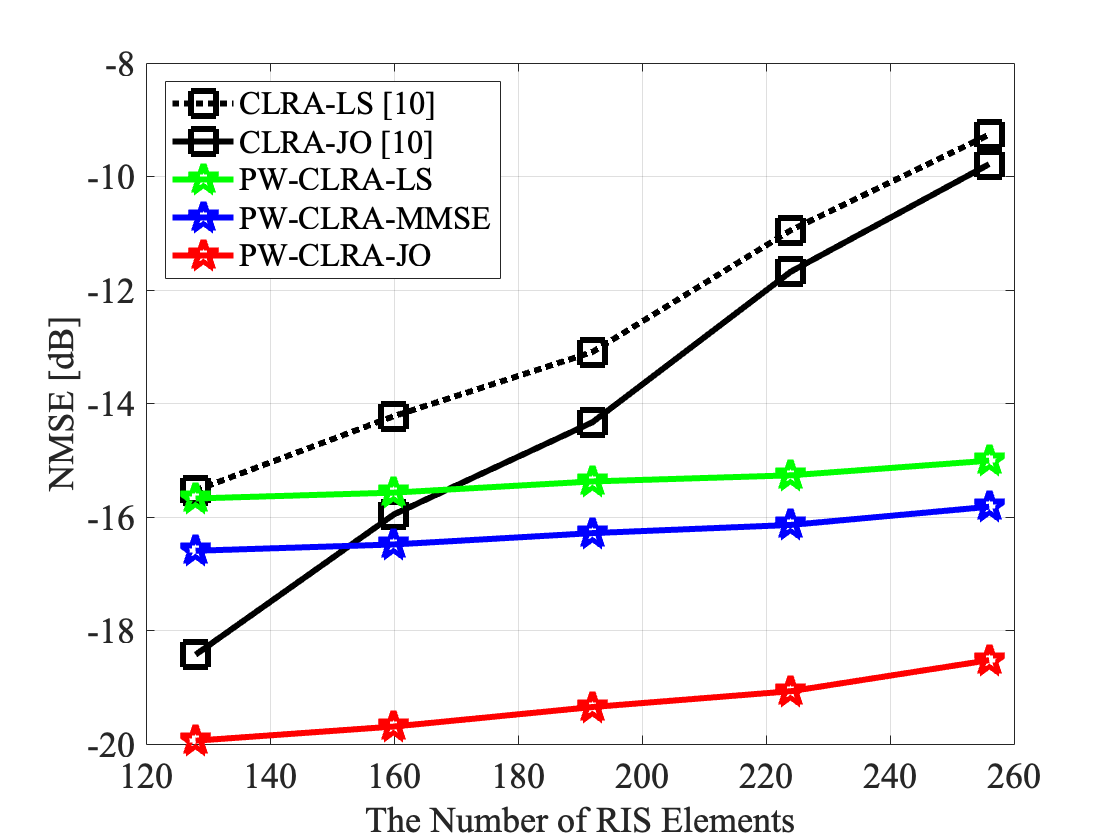}
\caption{The NMSE on the number of RIS reflection elements. $M\in\{128, 160, 192, 224, 256\}$, $K=B_c=Q=4$, $B_r = 1$, ${\rm SNR}=10\;{\rm dB}$ and $d_{x}^{\rm RB}=50\;{\rm m}$.}
\end{figure}

\begin{figure}[t]
\centering
\includegraphics[width=1.0\linewidth]{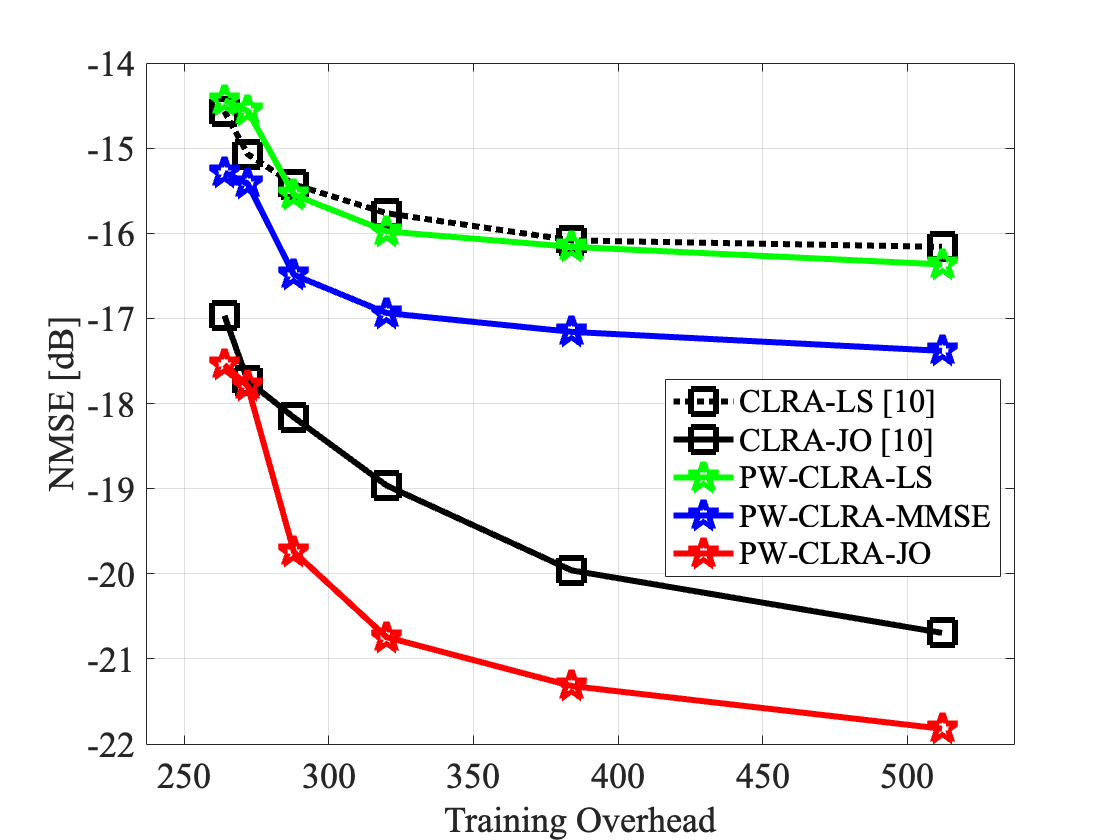}
\caption{The NMSE on the training overhead for UPA systems. $K=4$, $M=128\times 2=256$, ${\rm SNR}=10\;{\rm dB}$ and $d_{x}^{\rm RB} = 50\;{\rm m}$.}
\end{figure}

Fig. 7 depicts the NMSE on the number of RIS elements, in which the number of subsurface is fixed as $Q=4$. From \eqref{eq:np}, we can see that the noise power can be reduced as the number of RIS elements grows. On the contrary, it can possibly increase the rank of the submatrix (i.e., ${\rm rk}\left(\Hm_q^{\rm RB}\right)$), which might degrade the performance of the piece-wise column space estimation. From Fig. 7, we figure out that the PW-CLRA methods provide stable performances over the wide range of the number of RIS elements, while having the small number of subsurfaces (e.g., $Q=4$). However, the CLRA methods fail to handle the high-rank channels and thus, their performance severely get worse when the number of RIS elements is large.

Beyond the simple ULA, a uniform planar array (UPA) is widely used for practicality in 3-D coordinate systems \cite{Yang2024, Lee2024near}. Motivated by this, we conduct an additional experiment to verify that the proposed PW-CLRA methods can be straightforwardly applied to the UPA. Fig. 8 depicts the NMSE on the training overhead for UPA systems. We first observe that the proposed PW-CLRA-JO achieves a notable performance while having a small number of training overhead, which indeed verifies that the proposed method can be directly applied to the UPA systems without any modification. From Fig. 4 and Fig. 8, we can see that the estimation accuracy in the UPA systems is better than the ULA counterpart. In particular, the estimation accuracy of the CLRA-based methods is significantly enhanced showing almost same accuracy with $M=128$ case in Fig. 7. This is due to the following fact that the antenna aperture of UPA, determined by the diagonal length of the planar array, is much shorter than ULA in the fixed number of RIS elements, which can shorten the Rayleigh distance and thus reduce the rank of the RIS-BS channel. For example, the antenna aperture of the ULA with $256$ RIS elements is $256\lambda/2$ while that of a rectangular UPA with $128\times2 = 256$ RIS elements is $\sqrt{128^2 + 2^2}\lambda\approx 128\lambda/2$. Moreover, the antenna aperture of the square planar array (i.e., $16\times16=256$) becomes much shorter than the thin-rectangular planar array. That is, the near-field effect becomes negligible in the UPA systems due to the shorter Rayleigh distance. Nonetheless, in future wireless communication systems, the number of RIS elements would be extremely enlarged to the thousands for achieving higher data rates, in which the near-field channel should be considered as the Rayleigh distance also increases for UPA systems. Therefore, we can conclude that the proposed method would be a good candidate for extremely large-scale RIS-assisted MU-MIMO systems due to its attractive performance, lower training overhead and computational complexity, and the robustness to the large number of RIS elements.

\section{Conclusion}

We investigated the mixed LoS/NLoS near-field channel estimation for RIS-assisted MU-MIMO systems with hybrid beamforming structures. In these system, we for the first time proposed an efficient channel estimation method by overcoming the challenges of high-rankness and non-sparsity of the near-field LoS channels. The main idea comes from a piece-wise low-rank channel decomposition. Also, we designed novel reflection vectors and RF beamforming matrices suitable for the proposed piece-wise estimation. Via simulations, it was demonstrated that the proposed method can attain multi-user gains and yield a notable channel estimation accuracy while having a lower training overhead. Our on-going work is to pursue the beamforming optimization based on our estimated effective channel, focusing on the beam split (or squint) phenomenon emerging in near-field wideband communication systems.



\end{document}

%% file: macros.tex
\long\def\comment#1{}

\newfont{\bbb}{msbm10 scaled 700}

\newfont{\bb}{msbm10 scaled 1100}
\newcommand{\CC}{\mbox{\bb C}}

\newcommand{\EE}{\mbox{\bb E}}

\newcommand{\av}{{\bf a}}
\newcommand{\bv}{{\bf b}}

\newcommand{\tv}{{\bf t}}
\newcommand{\uv}{{\bf u}}
\newcommand{\wv}{{\bf w}}
\newcommand{\vv}{{\bf v}}
\newcommand{\xv}{{\bf x}}

\newcommand{\Am}{{\bf A}}

\newcommand{\Cm}{{\bf C}}
\newcommand{\Dm}{{\bf D}}

\newcommand{\Fm}{{\bf F}}

\newcommand{\Hm}{{\bf H}}
\newcommand{\Id}{{\bf I}}

\newcommand{\Mm}{{\bf M}}
\newcommand{\Nm}{{\bf N}}

\newcommand{\Pm}{{\bf P}}

\newcommand{\Sm}{{\bf S}}
\newcommand{\Tm}{{\bf T}}
\newcommand{\Um}{{\bf U}}
\newcommand{\Wm}{{\bf W}}
\newcommand{\Vm}{{\bf V}}
\newcommand{\Xm}{{\bf X}}
\newcommand{\Ym}{{\bf Y}}

\newcommand{\Ic}{{\cal I}}

\newcommand{\Mc}{{\cal M}}
\newcommand{\Nc}{{\cal N}}
\newcommand{\Oc}{{\cal O}}
\newcommand{\Pc}{{\cal P}}

\newcommand{\Sigmam}{\hbox{\boldmath$\Sigma$}}
\newcommand{\Phim}{\hbox{\boldmath$\Phi$}}

\newcommand{\diag}{{\hbox{diag}}}

\newcommand{\SNR}{{\sf SNR}}

\newcommand{\eqdef}{\stackrel{\Delta}{=}}

\newcommand{\herm}{{\sf H}}

\newcommand{\RED}{\color[rgb]{1.00,0.10,0.10}}
\newcommand{\BLUE}{\color[rgb]{0,0,0.90}}